\documentclass[aps,prd,onecolumn,superscriptaddress,showpacs]{revtex4}

\usepackage{graphicx}
\usepackage{amsmath}

\begin{document}

\title{Discrete and continuous symmetries in multi-Higgs-doublet models}

\author{P.\ M.\ Ferreira}
\affiliation{Instituto Superior de Engenharia de Lisboa,
    Rua Conselheiro Em\'{\i}dio Navarro,
    1900 Lisboa, Portugal}
\affiliation{Centro de F\'{\i}sica Te\'{o}rica e Computacional,
    Faculdade de Ci\^{e}ncias,
    Universidade de Lisboa,
    Av.\ Prof.\ Gama Pinto 2,
    1649-003 Lisboa, Portugal}
\author{Jo\~{a}o P.\ Silva}
\affiliation{Instituto Superior de Engenharia de Lisboa,
    Rua Conselheiro Em\'{\i}dio Navarro,
    1900 Lisboa, Portugal}
\affiliation{Centro de F\'{\i}sica Te\'{o}rica de Part\'{\i}culas,
    Instituto Superior T\'{e}cnico,
    P-1049-001 Lisboa, Portugal}

\date{\today}

\begin{abstract}
We consider the Higgs sector of multi-Higgs-doublet
models in the presence of simple symmetries
relating the various fields.
We construct basis invariant observables which 
may in principle be used to detect these symmetries
for any number of doublets.
A categorization of the symmetries into classes
is required,
which we perform in detail for the case of
two and three Higgs doublets.
\end{abstract}


\maketitle

\section{\label{sec:intro}Introduction}

Many features of the Standard Model (SM) of electroweak interactions
have been accurately tested.
Still, the Higgs sector remains largely unknown.
Indeed,
even after one scalar particle is directly detected,
there may well be further scalars awaiting discovery
(as required, for example, by supersymmetry).
It is easy to construct a $N$-Higgs-doublet model (NHDM),
but the number of parameters in the Higgs potential
grows very rapidly with $N$.
A generic two-Higgs-doublet model (THDM)
has 14 real parameters in the
Higgs potential,
while the generic three-Higgs-doublet model (3HDM)
already has 54 real parameters.

The number of parameters will be reduced if the
theory has discrete (or continuous) symmetries
relating the various Higgs fields,
which we denote by Higgs Family symmetries or
HF-symmetries.
Besides parameter reduction,
such symmetries may also be desirable features of
a theory in order to preclude flavor changing neutral currents
or to explain relations among different observables.
This article presents some features of
HF-symmetries in the NHDM.

Two deceptively simple questions about
HF-symmetries arise.
First,
classifying the symmetries by their impact
on the Higgs potential,
one would like to know how many distinct classes
of symmetries may be implemented.
Surprisingly,
this turns out to be a rather nontrivial question.
Second,
a simple basis change among the various Higgs fields
alters the Lagrangian but, obviously,
not its physical consequences.
Signals of HF-symmetries should be invariant
under these transformations.

The need to seek basis invariant observables in models with
many Higgs was pointed out by Lavoura and Silva
\cite{LS},
and by Botella and Silva \cite{BS},
stressing applications to CP violation.
Refs.~\cite{BS,BLS} indicate how to construct
basis invariant quantities in a systematic fashion
for any model,
including multi-Higgs-doublet models.
Work on basis invariance in the THDM was
much expanded upon by
Davidson and Haber \cite{DavHab},
by Gunion and Haber \cite{GunHab,Gun},
and by Haber and O'Neil \cite{HabONe}.
Basis invariance in the THDM was also considered in
Refs.~\cite{Bra,GinKra,Nishi,Ivanov}.
In particular,
Davidson and Haber \cite{DavHab}
develop several strategies to construct
basis invariant descriptions of HF-symmetries,
in the context of the THDM \cite{Lav}.
One of our aims is to extend their work 
into multi-Higgs systems.

The paper is organized as follows.
In section~\ref{sec:N-Higgs} we
introduce our notation and show that a simple HF-symmetry may
always be reduced to a standard diagonal form through a basis
transformation.
Then, we turn to the problem of classifying the
HF-symmetries according to their action on the
Higgs potential.
We cover the THDM in section~\ref{sec:N=2},
and we discuss the 3HDM in section~\ref{sec:N=3}.
In section~\ref{sec:fingerprints} we define a set of
basis invariant observables applicable to any NHDM,
which may in principle be used in order to identify
the presence of HF-symmetries.
We present our conclusions in section~\ref{sec:conclusions}.
Appendix~\ref{app:matrices} includes the
implications that the different classes of symmetries
of the 3HDM have on the quadratic and quartic coupling
coefficients.
This provides the fingerprint database against which
the basis invariant observables of section~\ref{sec:fingerprints}
should be compared.

\section{\label{sec:N-Higgs}The scalar sector of a generic
N-Higgs-doublet model}

\subsection{The scalar potential and basis transformations}

In this article we follow the notation of Refs.~\cite{BS,BLS,DavHab}.
Let us consider a $SU(2) \otimes U(1)$ gauge theory with
$N$ Higgs-doublets $\Phi_i$,
with the same hypercharge $1/2$,
and with vacuum expectation values (vevs)
\begin{equation}
\langle \Phi_i \rangle
=
\left(
\begin{array}{c}
0\\
v_i/\sqrt{2}.
\end{array}
\right)
\label{vev}
\end{equation}
The index $i$ runs from $1$ to $N$,
and we use the standard definition for the electric
charge, 
whereby the upper components of the $SU(2)$ doublets are
charged and the lower components neutral.

The scalar potential may be written as
\begin{equation}
V_H = Y_{ij} (\Phi_i^\dagger \Phi_j) +
Z_{ij,kl} (\Phi_i^\dagger \Phi_j) (\Phi_k^\dagger \Phi_l),
\label{Higgs_potential}
\end{equation}
where Hermiticity implies
\begin{eqnarray}
Y_{ij} &=& Y_{ji}^\ast,
\nonumber\\
Z_{ij,kl} \equiv Z_{kl,ij} &=& Z_{ji,lk}^\ast.
\label{hermiticity_coefficients}
\end{eqnarray}
The number of independent parameters of this potential is shown in
Table~\ref{parameters}.

\begin{table}[ht]
\caption{Number of parameters in the $Y$ and $Z$ coefficients
of the Higgs potential.}
\begin{ruledtabular}
\begin{tabular}{cccc}
& parameters & magnitudes  &  phases \\
\hline
$Y$ & $N^2$ & $\frac{N(N+1)}{2}$ & $\frac{N(N-1)}{2}$ \\
$Z$ &  $\frac{N^2(N^2+1)}{2}$ 
             & $\frac{N^2(N^2+3)}{4}$ & $\frac{N^2(N^2-1)}{4}$ \\
$Y$ and $Z$ & $\frac{N^2(N^2+3)}{2}$
             & $\frac{N^4+5N^2+2N}{4}$ & $\frac{N^4+N^2-2N}{4}$ \\
\end{tabular}
\end{ruledtabular}
\label{parameters}
\end{table}

The stationarity conditions are
\begin{equation}
\left[ Y_{ij}
+ 2 Z_{ij,kl}\, v_k^\ast v_l \right]\ v_j = 0
\hspace{3cm}(\textrm{for\ } i = 1, \cdots, N).
\label{stationarity_conditions}
\end{equation}
Multiplying by $v_i^\ast$ leads to
\begin{equation}
Y_{ij} (v_i^\ast \nu_j) = - 2 Z_{ij,kl}\,
(v_i^\ast v_j)\, (v_k^\ast v_l).
\label{aux_1}
\end{equation}

We may rewrite the potential in terms of new fields $\Phi^\prime_i$,
obtained from the original ones by a simple basis transformation
\begin{equation}
\Phi_i \rightarrow \Phi_i^\prime = U_{ij} \Phi_j,
\label{basis-transf}
\end{equation}
where $U$ is a $N \times N$ unitary matrix.
Under this unitary basis transformation,
the gauge-kinetic terms remain the same but the
coefficients $Y_{ij}$ and $Z_{ij,kl}$ are transformed as
\begin{eqnarray}
Y_{ij} & \rightarrow & 
Y^\prime_{ij} = 
U_{ik}\, Y_{kl}\, U_{jl}^\ast ,
\label{Y-transf}
\\
Z_{ij,kl} & \rightarrow & 
Z^\prime_{ij,kl} =
U_{im}\, U_{ko}\, Z_{mn,op}\, U_{jn}^\ast \, U_{lp}^\ast ,
\label{Z-transf}
\end{eqnarray}
and the vevs are transformed as
\begin{equation}
v_i \rightarrow v_i^\prime = U_{ij} v_j,
\label{vev-transf}
\end{equation}
Thus,
the basis transformations $U$ may be utilized in order to absorb
some of the parameters in $Y$ and/or $Z$,
meaning that not all parameters in
Table~\ref{parameters} have physical significance.

\subsection{\label{subsec:HFsymmetry}Higgs Family symmetries}

Let us assume that the scalar potential in
Eq.~(\ref{Higgs_potential}) has some explicit internal symmetry.
That is,
we assume that the coefficients of $V_H$ stay
\textit{exactly the same} under a transformation
\begin{equation}
\Phi_i \rightarrow \Phi_i^S = S_{ij} \Phi_j.
\label{S-transf-symmetry}
\end{equation}
$S$ is a unitary matrix,
so that the gauge-kinetic couplings
are also left invariant by this HF-symmetry.
As a result of this symmetry
\begin{eqnarray}
Y_{ij} & = & 
Y^S_{ij} = 
S_{ik}\, Y_{kl}\, S_{jl}^\ast ,
\label{Y-S}
\\
Z_{ij,kl} & = &
Z^S_{ij,kl} =
S_{im}\, S_{ko}\, Z_{mn,op}\, S_{jn}^\ast \, S_{lp}^\ast ,
\label{Z-S}
\end{eqnarray}
Notice that this is \textit{not} the situation considered
in Eqs.~(\ref{basis-transf})--(\ref{Z-transf}).
There,
the coefficients of the Lagrangian 
\textit{do change}.
What we said there was that,
although the coefficients do change,
the quantities which are physically measurable cannot.
What we consider in Eqs.~(\ref{S-transf-symmetry})--(\ref{Z-S})
is different.
Here we consider the possibility that
$V_H$ has some HF-symmetry $S$
which leaves the coefficients unchanged.

We now turn to the complicated interplay between
HF-symmetries and basis transformations.
Let us imagine that,
when written in the basis of fields $\Phi_i$,
$V_H$ has a symmetry $S$.
Then we perform a basis transformation from
the basis $\Phi_i$ to the basis $\Phi^\prime_i$,
as given by Eq.~(\ref{basis-transf}).
Clearly,
when written in the new basis,
$V_H$ does \textit{not} remain invariant under $S$.
Rather, it will be invariant under
\begin{equation}
S^\prime = U S U^\dagger .
\label{S-prime}
\end{equation}
As we change basis,
the form of the potential changes
in a way which may obscure the presence
of a HF-symmetry.
Eq.~(\ref{S-prime}) means that many HF-symmetries
which might look distinct on the surface,
will actually imply exactly the same physical predictions.
Any two symmetries $S$ and $S^\prime$ related by
Eq.~(\ref{S-prime}),
for some basis transformation $U$,
will make the same predictions.
Now $S$, $S^\prime$, and $U$ are all matrices of the $U(N)$ group,
within which Eq.~(\ref{S-prime}) constitutes a conjugacy relation.
Thus,
Eq.~(\ref{S-prime}) means that HF-symmetries
associated with matrices $S$ and $S^\prime$
in the same conjugacy class of $U(N)$ correspond to the same model.
This result is easy to generalize because an overall phase
transformation on $U$ or $S$ has no impact on the
potential $V_H$.
This can be seen directly from
Eqs.~(\ref{Y-transf})--(\ref{Z-transf})
and
Eqs.~(\ref{Y-S})--(\ref{Z-S}),
and is due to the fact that the Higgs potential
$V_H$ in Eq.~(\ref{Higgs_potential}) only depends
on the Higgs fields through bilinear combinations.
So, symmetries $S$ and $S^\prime$ belonging to conjugacy
classes related by a global phase transformation
lead to the same physics.

\subsection{\label{subsec:SpecialBasis}A special basis}

One can show that a $N \times N$ complex matrix $S$ belongs to
$U(N)$ if and only if there exists a unitary matrix $U$
such that $S^\prime$ in Eq.~(\ref{S-prime}) is diagonal,
with all entries of magnitude 1 \cite{Owen}.
This means that,
by a suitable basis transformation,
any symmetry $S$ may be brought to the form
\begin{equation}
\left(
\begin{array}{cccc}
e^{i \theta_1} & & & \\
 & e^{i \theta_2}& & \\
 & & \ddots & \\
& & & e^{i \theta_N} 
\end{array}
\right),
\label{S-diagonal-1}
\end{equation}
where $0 \leq \theta_i < 2\pi$ ($i = 1 \dots N$).
The conjugacy classes can thus be classified
by matrices of the type in Eq.~(\ref{S-diagonal-1}).

In the basis where the symmetry is
represented by Eq.~(\ref{S-diagonal-1}),
the coefficients must obey
\begin{eqnarray}
Y_{ij} & = & 
e^{i(\theta_i - \theta_j)} \, Y_{ij},
\label{Y-S-diag}
\\
Z_{ij,kl} & = &
e^{i(\theta_i - \theta_j)}\,
e^{i(\theta_k - \theta_l)}\,
Z_{ij,kl},
\label{Z-S-diag}
\end{eqnarray}
where there is no sum over repeated indexes.
This result is obtained by substituting the special form of
$S$ in Eq.~(\ref{S-diagonal-1}) onto Eqs.~(\ref{Y-S})--(\ref{Z-S}).
The result in Eq.~(\ref{Y-S-diag}) applies not only to the matrix $Y$
but to any matrix whose two indexes transform as $U^\dagger$
and $U$,
as shown for $Y$ in Eq.~(\ref{Y-transf}).
In particular,
in this special basis,
the matrices
\begin{eqnarray}
Z^{(1)}_{ij} & = &
\sum_k Z_{ik,kj},
\label{Z1}\\
Z^{(2)}_{ij} & = &
\sum_k Z_{ij,kk},
\label{Z2}
\end{eqnarray}
introduced by Davidson and Haber \cite{DavHab} must
also obey a relation like Eq.~(\ref{Y-S-diag}).
(Although sum over repeated indexes is assumed unless
explicitly stated,
we have shown it here explicitly for clarity.)

For symmetries corresponding to $\theta_i \neq \theta_j$
for all $i \neq j$,
Eq.~(\ref{Y-S-diag}) implies that the matrix $Y$ is diagonal.
If the Higgs potential only had quadratic terms,
this information would be useless,
since any hermitian matrix can be diagonalized by a suitable
unitary basis change.
Said otherwise,
without quartic terms in the potential,
imposing HF-symmetries (or not)
would make no difference.
Thus,
any sign of HF-symmetries must necessarily involve
the quartic terms.
For the special case of the THDM,
this can be seen explicitly in Eqs.~(39)--(50)
of Ref.~\cite{DavHab}

But given two matrices
(for example, $A=Y$ and $B=Z^{(1)}$),
Eq.~(\ref{Y-S-diag}) already gives crucial information
in the case where
$\theta_i \neq \theta_j$ for all $i \neq j$.
Indeed it states that,
in the special basis,
$A$ and $B$ are simultaneously diagonal.
As a result,
their commutator vanishes;
$\left[ A, B\right]=0$.
Now, the commutator is a matrix and the null matrix
is always mapped onto the null matrix,
regardless of which basis transformation one chooses.
Thus,
we conclude that symmetries which are represented in the
special basis by $\theta_i \neq \theta_j$
(for all $i \neq j$) will lead to the basis-invariant
result $\left[ A, B\right]=0$.
This can be used to define basis invariant fingerprints
of HF-symmetries,
explaining Eqs.~(39)--(41) of Ref.~\cite{DavHab}

One could now ask whether all possible impositions
due to HF-symmetries can be cast in the form
$\left[ A, B\right]=0$ for suitably chosen
matrices $A$ and $B$.
The answer is negative,
as Davidson and Haber found when trying to
disentangle the Peccei-Quinn \cite{PQ}
symmetry from the
usual $Z_2$ symmetry \cite{Pas}
-- \textit{c.f.} their Eq.~(46).

\subsection{\label{subsec:Global}Further simplifications due
to global phase invariance}

Because the Lagrangian is invariant under global
phase transformations,
there are infinitely many conjugacy classes which
imply the same physical predictions.
Indeed,
classes represented by the diagonal elements
\begin{equation}
e^{i \theta_1} 
\left(
\begin{array}{cccc}
\: 1 \: & & & \\
 & e^{i \theta_2}& & \\
 & & \ddots & \\
& & & e^{i \theta_N} 
\end{array}
\right)
\label{S-diagonal-2}
\end{equation}
for fixed values of $\theta_j$ ($j=2 \dots N$) lead to the same
physical predictions,
regardless of the value of $\theta_1$.
This means that we can concentrate on symmetries of the type
shown in Eq.~(\ref{S-diagonal-2}),
without the pre-factor $e^{i \theta_1}$.
Thenceforth,
we shall classify each class of symmetries by their
diagonal representative:
\begin{equation}
S = \left(
\begin{array}{cccc}
\: 1 \: & & & \\
 & e^{i \theta_2}& & \\
 & & \ddots & \\
& & & e^{i \theta_N} 
\end{array}
\right)
\label{S-diagonal-3}
\end{equation}
Alternatively,
we could use the $e^{i \theta_1}$ phase freedom in order
to restrict our attention to symmetries in $SU(N)$.

A first question now arises: do classes corresponding to
different values of $\theta_j$ ($j=2 \dots N$) necessarily
imply different physical predictions?
The answer is negative.
A second question arises:
can one classify the different types of HF-symmetries
according to their impact on the Higgs potential?
The answer is affirmative,
but that must be done separately for each value of $N$.
We review the case of $N=2$ in
Section~\ref{sec:N=2},
and we turn to the more difficult case of
$N=3$ in Section~\ref{sec:N=3}.

\section{\label{sec:N=2}HF-Symmetries in the THDM}

In the previous sections we learned the following.
Two symmetries in the same conjugacy class yield the same
physics.
Thus,
we can go into a special basis and consider
a diagonal matrix with complex entries of unit magnitude.
In fact,
there are infinitely many such diagonal matrices
which yield the same physics,
because global phase transformations have no
impact on the Lagrangian.
Therefore, we can concentrate on symmetries of the type
\begin{equation}
S=
\left(
\begin{array}{cc}
\: 1 \: & \\
 & e^{i \alpha}\\
\end{array}
\right).
\label{S-2x2}
\end{equation}
Substituting into Eq.~(\ref{Y-S-diag}),
we obtain
\begin{equation}
\left(
\begin{array}{cc}
  Y_{11} & Y_{12}\\
  Y_{21} & Y_{22}
\end{array}
\right)
=
\left(
\begin{array}{cc}
  Y_{11} & e^{-i \alpha} Y_{12}\\
  e^{i \alpha} Y_{21} & Y_{22}
\end{array}
\right)
\label{Y_with_phases}
\end{equation}
We conclude that the $Y$ matrix elements come
affected by the following phase factors:
\begin{equation}
\left[
\begin{array}{cc}
0 & - \alpha\\
 \alpha & 0 \\
\end{array}
\right].
\label{SU2-Y}
\end{equation}
This table is a shorthand notation to keep track of the
exponents which appear in Eq.~(\ref{Y_with_phases}).
If $S$ is indeed a symmetry of the potential,
then these phase factors must equal 0 (mod $2\pi$)
for any nonzero value of the corresponding entry
in the $Y_{ij}$ matrix.
If $\alpha=0$,
we have the uninteresting identity transformation.
We shall ignore this possibility henceforth.
If $ 0 < \alpha < 2\pi$,
then any matrix in the problem
(built from the coefficients in the scalar potential)
must be diagonal.
This leads to conditions of the type
$\left[ A, B\right]=0$ discussed above.
Notice that this condition does not distinguish
$\alpha=\pi$,
corresponding to the $Z_2$ symmetry
\begin{equation}
S_1=
\left(
\begin{array}{cc}
1 & 0 \\
0 & -1 \\
\end{array}
\right),
\label{Z_2}
\end{equation}
from the transformation with $\alpha=\pi/3$, etc.
Accordingly,
Davidson and Haber \cite{DavHab} were unable to find a condition
to distinguish $Z_2$ from Peccei-Quinn based exclusively
on matrix conditions of the type $\left[ A, B\right]=0$.

The $Z_2$ symmetry will only be distinguished from the symmetries
with other values of $\alpha$ by the quartic terms.
Substituting Eq.~(\ref{S-2x2}) into Eq.~(\ref{Z-S-diag}),
we conclude that the $Z$ matrix elements come
affected by the following phase factors:
\begin{equation}
\left[
\begin{array}{cc}
    \left[
    \begin{array}{cc}
      0 & - \alpha\\
      \alpha & 0 \\
    \end{array}
    \right]
&
    \left[
    \begin{array}{cc}
      - \alpha & - 2\alpha\\
      0 & - \alpha \\
    \end{array}
    \right]
\\*[7mm]
    \left[
    \begin{array}{cc}
      \alpha & 0 \\
      2\alpha & \alpha \\
    \end{array}
    \right]
&
    \left[
    \begin{array}{cc}
      0 & - \alpha\\
      \alpha & 0 \\
    \end{array}
    \right]
\\
\end{array}
\right].
\label{SU2-Z}
\end{equation}
This is represented as a table of tables.
The uppermost-leftmost table corresponds to
the phases affecting $Z_{11,kl}$. 
The next table along the same line corresponds
to the phases affecting $Z_{12,kl}$, and so on...
If $S$ is indeed a symmetry of the potential,
then these phase factors must equal 0 (mod $2\pi$)
for any nonzero value in the corresponding entry
of the $Z_{ij,kl}$ tensor.
Unlike what happened for the quadratic terms
(and, in general, for any matrix built out of
quadratic and/or quartic terms)
we see that there is a distinction between two
cases,
according to whether
$2\alpha=2\pi$ or $2\alpha \neq 0, 2\pi$.
If $\alpha=\pi$,
then the terms $Z_{12,12}$ and $Z_{21,21}$
(which are related to a parameter denoted by $\lambda_5$
in usual presentations of the THDM)
may be different from zero.
In contrast,
$\lambda_5$ must vanish
for symmetries with $\alpha \neq 0,\pi$.

So,
the quartic terms do distinguish $S_1$ in Eq.~(\ref{Z_2})
from
\begin{equation}
S_2=
\left(
\begin{array}{cc}
1 & 0 \\
0 & e^{i \alpha} \\
\end{array}
\right)_{(\alpha \neq 0, \pi)}.
\label{PQ_1}
\end{equation}
But they do not distinguish among the symmetries
\begin{equation}
S_{2/3} =
\left(
\begin{array}{cc}
1 & 0 \\
0 & e^{i 2\pi/3} \\
\end{array}
\right),
\hspace{3ex}
S_{2/5} =
\left(
\begin{array}{cc}
1 & 0 \\
0 & e^{i 2\pi/5} \\
\end{array}
\right).
\hspace{3ex}
\end{equation}
These symmetries are actually quite curious.
Suppose we impose the symmetry $S_{2/3}$ on the Lagrangian.
Clearly,
applying the symmetry again must also leave the Lagrangian invariant.
As a result, the Lagrangian is invariant under
$S_{2/3}$, $S_{2/3}^2$, and $S_{2/3}^3=1$,
which form a closed group.
The Lagrangian is always invariant under a group;
if it is invariant under symmetries $S_a$ and $S_b$,
it is obviously also invariant under $S_a S_b$.
And if we choose $\alpha/\pi$ irrational,
the group $S_2, S_2^2, S_2^3, \dots$
will even have an infinite number of elements.

But there is a further important point.
We have just shown that if the potential is invariant
with respect to a symmetry $S_2$ for some value
of $\alpha \neq 0, \pi$,
then it will necessarily be invariant with respect to a
symmetry $S_2$ with any other value of $\alpha$.
That is,
we have imposed a discrete symmetry but the resulting potential
is invariant with respect to a continuous symmetry --
the Peccei-Quinn symmetry \cite{PQ}.
This is an important point because continuous symmetries,
if broken,
imply the presence of massless Goldstone bosons.
Suppose we build a NHDM with an innocent-looking
discrete symmetry.
It may happen that imposing this symmetry has the
same effect on the potential as a global symmetry
and, thus,
the possibility exists for undesired massless scalars.
We have just seen one such example.
We impose the discrete symmetry $S_{2/3}$
(the corresponding group of symmetries,
to be precise)
only to find that the resulting potential is
invariant under the continuous Peccei-Quinn symmetry.

Notice that the analysis of the quartic terms is sufficient
to isolate all cases of interest.
Indeed,
the uppermost-leftmost $2 \times 2$ block of
Eq.~(\ref{SU2-Z}) coincides with Eq.~(\ref{SU2-Y}).
A similar situation occurs for any other value of $N$.

In conclusion, as far as simple HF-symmetries are concerned,
we have only three possibilities.
Either we have the most general Lagrangian,
or we have the $Z_2$ symmetry,
or we have the PQ symmetry.
This exhausts all simple symmetries.
We are not considering here CP-type symmetries,
and we comment briefly on multiple symmetries in
section~\ref{subsec:multiple}.

\subsection{\label{subsec:multiple}Multiple symmetries}

In this article we concentrate on what we call
simple symmetries.
By this we mean the following:
we choose some symmetry $S$ and we impose only that
symmetry on the Higgs potential.

We recall two points.
First,
under a basis change, that symmetry will look different.
For example,
if we impose the symmetry $\Phi_1 \rightarrow \Phi_1$,
$\Phi_2 \rightarrow - \Phi_2$ on some basis,
then that symmetry will turn into
$\Phi^\prime_1 \leftrightarrow \Phi^\prime_2$
if we change into the basis
$\Phi^\prime_1 =(\Phi_1 + \Phi_2)/\sqrt{2}$,
$\Phi^\prime_2 =(\Phi_1 - \Phi_2)/\sqrt{2}$.
Indeed,
\begin{equation}
D =
\left(
\begin{array}{cc}
0 & 1 \\
1 & 0 \\
\end{array}
\right)
=
\frac{1}{\sqrt{2}}
\left(
\begin{array}{cc}
1 & 1 \\
1 & -1 \\
\end{array}
\right)
\ 
\left(
\begin{array}{cc}
1 & 0 \\
0 & -1 \\
\end{array}
\right)
\ 
\frac{1}{\sqrt{2}}
\left(
\begin{array}{cc}
1 & 1 \\
1 & -1 \\
\end{array}
\right)
\label{D}
\end{equation}
is the corresponding Eq.~(\ref{S-prime})

Second,
it may be that imposing that symmetry alone
will yield a potential with a larger symmetry.
We are not referring only to the obvious
possibility that the potential becomes automatically
invariant to the group of all powers of $S$
(as exemplified above in connections with $S_{2/3}$).
It may be that the potential becomes automatically invariant
under other symmetries,
such as continuous symmetries,
as indeed happens in the THDM.

All the possibilities discussed thus far fall
under what we call simple symmetries.
But we may have more complicated situations.
We may impose simultaneously two symmetries.
For example, we may ask that the potential be
invariant \textit{both} under $S_1$ in Eq.~(\ref{Z_2})
and $D$ in Eq.~(\ref{D}),
\textit{in the same basis}.
Notice that now it is irrelevant that $S_1$ and $D$ are
in the same conjugacy class when considered individually.
We are imposing both in the same basis.
Bringing one to diagonal form will make the other off-diagonal,
and vice-versa.
In this case,
the potential becomes automatically invariant under
the group of four symmetries $S_1$, $D$, $S_1 D$, and $1$.
This case is considered by Davidson and Haber \cite{DavHab}
after their Eq.~(37).
Other multiple symmetries of the THDM were studied by Ivanov
\cite{Ivanov}.
The general analysis of such cases in the NHDM is much more
difficult and it is not considered in this article.
Indeed,
what we dubbed simple symmetries will prove
surprisingly demanding,
even for $N=3$.

\section{\label{sec:N=3}HF-Symmetries in the 3HDM}

The analogues of Eqs.~(\ref{S-2x2}),
(\ref{SU2-Y}),
and (\ref{SU2-Z}),
for $N=3$ are

\begin{equation}
S=
\left(
\begin{array}{ccc}
\: 1 \: & & \\
 & e^{i \alpha} & \\
 & & e^{i \beta} \\
\end{array}
\right),
\label{S-3x3}
\end{equation}
\begin{equation}
\left[
\begin{array}{ccc}
0 & - \alpha & - \beta \\
\alpha & 0 &  \alpha - \beta \\
\beta & \beta - \alpha & 0 \\
\end{array}
\right],
\label{SU3-Y}
\end{equation}
and
\begin{equation}
\left[
\begin{array}{ccc}
    \left[
    \begin{array}{ccc}
      0 & - \alpha & - \beta \\
      \alpha & 0 &  \alpha - \beta \\
      \beta & \beta - \alpha & 0 \\
    \end{array}
    \right]
&
    \left[
    \begin{array}{ccc}
      - \alpha & - 2\alpha & - \alpha - \beta \\
      0 & - \alpha & - \beta \\
      \beta - \alpha & \beta - 2\alpha & - \alpha \\
    \end{array}
    \right]
&
    \left[
    \begin{array}{ccc}
      - \beta & - \alpha - \beta & - 2 \beta \\
      \alpha - \beta & - \beta & \alpha - 2\beta \\
      0 & - \alpha & - \beta \\
    \end{array}
    \right]
\\*[12mm]
    \left[
    \begin{array}{ccc}
      \alpha & 0 & \alpha - \beta \\
      2 \alpha & \alpha &  2\alpha - \beta \\
      \alpha + \beta & \beta & \alpha \\
    \end{array}
    \right]
&
    \left[
    \begin{array}{ccc}
      0 & - \alpha & - \beta \\
      \alpha & 0 &  \alpha - \beta \\
      \beta & \beta - \alpha & 0 \\
    \end{array}
    \right]
&
    \left[
    \begin{array}{ccc}
      \alpha - \beta & - \beta &  \alpha - 2 \beta \\
      2\alpha - \beta & \alpha - \beta & 2\alpha - 2\beta \\
      \alpha & 0 & \alpha - \beta \\
    \end{array}
    \right]
\\*[12mm]
    \left[
    \begin{array}{ccc}
      \beta & \beta - \alpha & 0 \\
      \alpha + \beta & \beta & \alpha \\
      2\beta & 2\beta - \alpha & \beta \\
    \end{array}
    \right]
&
    \left[
    \begin{array}{ccc}
      \beta - \alpha & \beta - 2\alpha & - \alpha \\
      \beta & \beta - \alpha & 0 \\
      2 \beta - \alpha & 2\beta - 2\alpha & \beta - \alpha \\
    \end{array}
    \right]
&
    \left[
    \begin{array}{ccc}
      0 & - \alpha & - \beta \\
      \alpha & 0 & \alpha - \beta\\
      \beta & \beta - \alpha & 0 \\
     \end{array}
    \right]
\\
\end{array}
\right],
\label{SU3-Z}
\end{equation}
respectively.

Let us first look at the impact of the symmetries on the
quadratic terms in Eq.~(\ref{SU3-Y}).
One interesting situation is
$\alpha = 0$, $\beta \neq 0$.
In this situation $\Phi_1$ and $\Phi_2$ have the same
transformation, while $\Phi_3$ transforms differently.
One could think of $\alpha \neq 0$, $\beta = 0$
as a different situation,
but it is not.
It is the same as the previous situation,
with the interchange of fields 2 and 3.
Since such a field permutation corresponds
to a basis change (achievable through
some unitary matrix $U$),
the two situations correspond to exactly the same
symmetry viewed in different basis,
and lead to the same physics.
Now we consider $\alpha \neq 0$, $\beta \neq 0$.
The $(2,3)$ and $(3,2)$ entries in Eq.~(\ref{SU3-Y}) show that we must
distinguish $\alpha = \beta$ from $\alpha \neq \beta$.
But $\alpha=\beta$ corresponds to the symmetry
\begin{equation}
\left(
\begin{array}{ccc}
\: 1 \: & & \\
 & e^{i \beta} & \\
 & & e^{i \beta} \\
\end{array}
\right)
=
e^{i \beta}
\left(
\begin{array}{ccc}
 e^{-i \beta} & & \\
 & \: 1 \:& \\
 & & \: 1 \:\\
\end{array}
\right),
\end{equation}
which, aside from the irrelevant overall phase,
is just a $1 \leftrightarrow 3$ permutation of the
situation already considered.
In conclusion,
the quadratic terms in the Higgs potential distinguish
among the following two types of symmetries:
\begin{equation}
\left(
\begin{array}{ccc}
\: 1 \: & & \\
 & 1 & \\
 & & e^{i \beta} \\
\end{array}
\right)_{(\beta \neq 0)}, \hspace{4ex}
\left(
\begin{array}{ccc}
\: 1 \: & & \\
 & e^{i \alpha} & \\
 & & e^{i \beta} \\
\end{array}
\right)_{(\alpha \neq 0 | \beta \neq \alpha, 0)}.
\label{two_from_Y}
\end{equation}

As happened for $N=2$,
looking at the quartic terms will further open up these classes
of symmetries.
Said otherwise,
two symmetries may have the same impact on the quadratic terms
but different impact on the quartic terms.
The zeros in Eq.~(\ref{SU3-Z}) correspond to the
entries of the $Z$ tensor which are real.
In order to see how the symmetries affect the quartic terms
we start by collecting all distinct combinations
in Eq.~(\ref{SU3-Z}):
$\alpha$, $2\alpha$, $\beta$, $2\beta$, $\alpha+\beta$,
$\alpha-\beta$, $2\alpha-\beta$, $\alpha - 2 \beta$,
and $2 \alpha - 2\beta$.
Each may be equal to 0 (mod $2\pi$) or not.
We study all possible combinations,
making sure that each new class of symmetries found
does not correspond to a mere basis transformation of a class
considered previously.
Proceeding in this fashion,
we find that
(as far as simple HF-symmetries are concerned,
and aside from the most general Lagrangian)
there are seven types of symmetries having
distinct impacts on the Higgs potential:
\begin{eqnarray}
& &
S_1 =
\left(
\begin{array}{ccc}
\: 1 \: & & \\
 & \: 1 \:& \\
 & & -1 \\
\end{array}
\right),
\hspace{4ex}
\hspace{25mm}
S_2 =
\left(
\begin{array}{ccc}
\: 1 \: & & \\
 & \: 1 \:& \\
 & & e^{i \alpha} \\
\end{array}
\right)_{(\alpha \neq 0, \pi)},
\nonumber\\*[4mm]
& &
S_3 =
\left(
\begin{array}{ccc}
\: 1 \: & & \\
 & e^{2i\pi/3} & \\
 & & e^{-2i\pi/3} \\
\end{array}
\right),
\hspace{4ex}
\hspace{14mm}
S_4 = 
\left(
\begin{array}{ccc}
\: 1 \: & & \\
 & \: i \:& \\
 & & -i \\
\end{array}
\right),
\nonumber\\*[4mm]
& & 
S_5 =
\left(
\begin{array}{ccc}
\: 1 \: & & \\
 & e^{i \alpha} & \\
 & & e^{-i \alpha} \\
\end{array}
\right)_{(\alpha \neq 0, \pi/2,  2\pi/3, \pi)},
\hspace{6mm}
S_6 =
\left(
\begin{array}{ccc}
\: 1 \: & & \\
 & - 1 & \\
 & & e^{i \alpha} \\
\end{array}
\right)_{(\alpha \neq 0, \pi)},
\nonumber\\*[4mm]
& &
S_7 =
\left(
\begin{array}{ccc}
\: 1 \: & & \\
 & e^{i \alpha} & \\
 & & e^{i \beta} \\
\end{array}
\right)_{(\alpha \neq 0, \pi | \beta \neq \pm \alpha, 0, \pi)}.
\label{seven_from_Z}
\end{eqnarray}
Comparing with Eq.~(\ref{two_from_Y}),
we see that $S_1$ and $S_2$ have the same impact
on the quadratic terms.
This is different from the impact of $S_3$--$S_7$ 
on the quadratic terms.
The impact of the symmetries $S_1$--$S_7$ on the
coefficients of the Higgs potential is presented
in appendix~\ref{app:matrices}.

Recall that the potential is always invariant under a group
of symmetries.
For example,
imposing $S_3$,
the potential is automaticaly invariant under
$S_3$, $S_3^2$, and $S_3^3=1$.
The symmetries $S_2$, $S_5$, and $S_6$ are
special in this respect.
Imagine that we impose $S_2$ from some value of
$\alpha \neq 0, \pi$.
Then,
the potential is automaticaly invariant with respect to
$S_2$ with any other value for $\alpha$.
That is,
we wish to impose a discrete symmetry,
but the resulting potential turns out to
be invariant under a $U(1)$ continuous symmetry.
The case for $S_7$ is even worse.
Imposing $S_7$ for some chosen numerical values
for $\alpha$ ($\neq 0, \pi$) and
$\beta$ ($\neq \pm \alpha, 0, \pi$)
will automaticaly generate a potential invariant
under all symmetries $S_7$ for any values of
$\alpha$ and $\beta$.
The resulting potential will be invariant under
a $U(1) \otimes U(1)$ continuous symmetry.

\section{\label{sec:fingerprints}Basis invariant descriptions of
the symmetry classes in the NHDM}

Let us look back at Eq.~(\ref{two_from_Y}).
We recall that the analysis of the quadratic terms applies
equally well to any matrix
(even if built out of the quartic terms,
as $Z^{(1)}$ and $Z^{(2)}$).
Therefore,
a (basis invariant) commutator type condition
will distinguish one symmetry of the first type
in Eq.~(\ref{two_from_Y})
($S_1$--$S_2$, where $\left[A,B\right] \neq 0$)
from one of the second type
($S_3$--$S_7$, where $\left[A,B\right]=0$).
But it will not distinguish among two symmetries
of the same type in Eq.~(\ref{two_from_Y}).
For example,
it will not distinguish $S_3$ from $S_7$,
even though they have a different impact on the quartic terms.
As a result,
commutator conditions,
which were so central to Davidson and Haber's
study of the THDM \cite{DavHab},
have a very limited use for $N \geq 3$.

Basis invariant fingerprints of the HF-symmetries
may be found by combining eigenvectors
of the matrix of quadratic couplings $Y$
with the quartic couplings $Z$.
In the context of the THDM this
seemed a curiosity and was left by Davidson
and Haber \cite{DavHab} to the end of their
appendix B; Eqs.~(B17)--(B22).
Inspired by this remark,
we developed a technique which,
when suitably extended and interpreted,
will become central to our definition of the basis
invariant observables identifying HF-symmetries.

Consider a general 3HDM.
We define the three eigenvectors of the matrix $Y$
by $\hat{y}^1$, $\hat{y}^2$, and $\hat{y}^3$.
If the matrix has two or three degenerate eigenvalues,
we are free to choose any orthonormal basis
in the degenerate space.
The components of the $\hat{y}^\alpha$ eigenvector
are denoted by $\hat{y}^\alpha_i$.
Under a basis change $U$,
the components of the eigenvectors change as
\begin{equation}
\hat{y}^\alpha_i \rightarrow U_{ij}\ \hat{y}^\alpha_j.
\end{equation}
Combining this with Eq.~(\ref{Z-transf}) we see that
the quantities
\begin{equation}
I^{\alpha \beta, \gamma \delta} \equiv
Z_{ij,kl}\, 
(\hat{y}^\alpha_i)^\ast \,
(\hat{y}^\beta_j) \,
(\hat{y}^\gamma_k)^\ast \,
(\hat{y}^\delta_l)
\label{crucial_1}
\end{equation}
are basis invariant for any values
of $\alpha$, $\beta$, $\gamma$, and $\delta$
between 1 and 3.
Therefore, we can evaluate them in any basis.
In particular,
in a basis where $Y$ is diagonal,
we may choose the eigenvectors as
\begin{equation}
\hat{y}^1
=
\left(
\begin{array}{c}
1\\
0\\
0
\end{array}
\right),  \hspace{7mm}
\hat{y}^2
=
\left(
\begin{array}{c}
0\\
1\\
0
\end{array}
\right), \hspace{7mm}
\hat{y}^3
=
\left(
\begin{array}{c}
0\\
0\\
1
\end{array}
\right).
\label{basis_Y_eigenvectors}
\end{equation}
That is, $\hat{y}^\alpha_i = \delta_{\alpha i}$,
where $\delta_{\alpha i}$ is the Kronecker symbol.
In the basis of Eq.~(\ref{basis_Y_eigenvectors}),
the quantities in Eq.~(\ref{crucial_1}) become
\begin{equation}
I^{\alpha \beta, \gamma \delta} \equiv
Z_{ij,kl}\, 
(\hat{y}^\alpha_i)^\ast \,
(\hat{y}^\beta_j) \,
(\hat{y}^\gamma_k)^\ast \,
(\hat{y}^\delta_l)
= Z_{\alpha \beta, \gamma \delta}.
\ \ \ \ \ \ 
\mbox{(mod permutations)}
\label{crucial_2}
\end{equation}
This means that the quantities $I^{\alpha \beta, \gamma \delta}$
(permutations aside) equal the quartic couplings 
$Z_{\alpha \beta, \gamma \delta}$ calculated in the basis
where $Y$ is diagonal.
As a result,
$I^{\alpha \beta, \gamma \delta}$ has the same symmetries of the
$Z$ couplings and,
according to Table~\ref{parameters},
only $N^2(N^2+1)/2$ of these are independent.

Before we proceed, we must point out a
subtlety concerning
Eqs.~(\ref{crucial_1})--(\ref{crucial_2}).
Suppose that we have a $Y$ matrix whose eigenvalues are not
degenerate,
and that we find its three eigenvectors.
Now we have a problem in attributing to them the labels
$1$, $2$, and $3$.
There are 6 possibilities,
which differ by permutations.
What one author chooses as $I^{1,1,1,2}$ may be what
another author chooses as $I^{2,2,2,1}$.
Once this choice is made,
then the quantity is basis invariant.
But the choices of different authors may differ by
permutations connected with their specific
choices for the ordering of the eigenvectors.
Eq.~(\ref{crucial_2}) identifies one possibility;
the other five possibilities differ by permutations in the
choice of basis eigenvectors in Eq.~(\ref{basis_Y_eigenvectors})
Thus,
permutations must be considered when using
appendix~\ref{app:matrices},
as explained below.

We now show how the quantities in Eq.~(\ref{crucial_1})
can be used to identify the various discrete symmetries.
Let us assume that the Higgs potential is invariant under 
some symmetry $S_3$--$S_7$; for example $S_3$.
The potential may be originally written in a basis
where this symmetry is not diagonal.
But that is irrelevant;
we may always consider what happens in a basis where
the symmetry has the form in Eq.~(\ref{seven_from_Z}).
In this basis the $Z$ coefficients have the structure
(of zero and non-zero entries) presented in
appendix~\ref{app:matrices} for the $S_3$ symmetry.
Also in this basis,
the $Y$ matrix is diagonal.
Therefore, its eigenvectors are given by
Eq.~(\ref{basis_Y_eigenvectors}) or some permutation
thereof.
We conclude from Eq.~(\ref{crucial_2})
that the observables in Eq.~(\ref{crucial_1})
must fall into the pattern shown in
appendix~\ref{app:matrices} which corresponds to $S_3$,
or some permutation thereof.

The discussion in the previous paragraph
invoked a special basis,
only to show that the presence of a symmetry
$S_3$--$S_7$ will force the observables of
Eq.~(\ref{crucial_1}) to fall onto the corresponding
pattern shown in appendix~\ref{app:matrices}
(or some permutation thereof).
But the invariant need not be calculated in this basis.
Because it is a basis invariant,
it can be calculated in any basis whatsoever;
the result must be the same.
So, the algorithm to identify the presence of
a symmetry is straightforward:
\begin{itemize}
\item We start with the potential in some original basis.
The potential has the symmetry $S_i$ in that basis
(in general $S_i$ will not have the simple diagonal form
when written in that original basis).
\item We find the eigenvectors of $Y$
(which, in general, will also not be diagonal
in the original basis).
\item We combine the $Y$ eigenvectors with $Z$
to calculate the basis invariant
$I^{\alpha \beta, \gamma \delta}$
observables in Eq.~(\ref{crucial_1}).
\item We check whether the resulting pattern
matches the patterns in appendix~\ref{app:matrices}
(or some permutation thereof).
\end{itemize}
This procedure identifies which symmetry we have,
even when the potential is written in an original basis
where $S_i$ has a very obscure form.

We have postponed the proof that this procedure also works
for $S_i$ ($i=1,2$) because the basis where
this symmetry is diagonal does not guarantee
that $Y$ is diagonal.
Indeed,
in a basis where $S_1$ (say) is diagonal,
$Y$ is block diagonal,
\textit{c.f.\/} appendix~\ref{app:matrices}.
In order to make $Y$ diagonal
and guarantee that its eigenvectors can be
cast in the form of Eq.~(\ref{basis_Y_eigenvectors})
(aside from permutations),
we need a further diagonalization of the
uppermost-leftmost $2\times2$ block of $Y$.
But because the two first eigenvalues
of $S_1$ and $S_2$ are degenerate,
a unitary $2\times2$ rotation on the
uppermost-leftmost block has no effect on
the form of the symmetry.
This shows that a basis may be found where
$S_1$--$S_2$ have the form in Eq.~(\ref{seven_from_Z})
and $Y$ is diagonal,
completing our proof.

Clearly,
Eqs.~(\ref{crucial_1}) and (\ref{crucial_2})
hold for any value of $N$.
As a result,
we have succeeded in defining basis invariant
quantities which can \textit{in principle} be utilized in order
to identify any HF-symmetry in NHDM,
for any value of $N$.
But in order to perform this
identification \textit{in practice} we need
to have a set of textures to compare with,
as we have done in appendix~\ref{app:matrices} for $N=3$.
The problem of categorizing the different classes
of HF-symmetries which may affect the Higgs potential
(and, thus, the corresponding textures) becomes
demanding as $N$ increases.
For example,
in a cursory analysis of $N=4$ we have identified
at least 15 distinct classes of symmetries.

One final remark concerns the possiblitity that
the matrix $Y$ has degenerate eigenvalues.
In this case we must define new
$I^{\alpha, \beta, \gamma, \delta}$ parameters
invoking the eigenvectors of $Z^{(1)}$ (or $Z^{(2)}$)
rather than the eigenvectors of $Y$.
Some regions of parameter space may require special care.
These types of questions were already present in the
various methods proposed for the THDM \cite{DavHab}.

An apt analogy to our procedure is the following.
We wish to identify a symmetry.
The $I^{\alpha \beta, \gamma \delta}$ in
Eq.~(\ref{crucial_1}) provide us with a
(basis independent) fingerprint of the symmetry.
But we can only use this information in order to identify
the symmetry,
if we have a database with all the
distinct fingerprints which may show up,
one for each symmetry class.
This is what we provide explicitly
in appendix~\ref{app:matrices} for $N=3$.
Anyone interested may construct a similar
database for $N \geq 4$.

\section{\label{sec:conclusions}Conclusions}

We have constructed a set of basis invariant
quantities which may in principle be used to identify
the presence of HF-symmetries in a NHDM,
regardless of the value of $N$.
HF-symmetries can be classified according to their impact
on the Higgs potential.
Surprisingly,
this classification is already involved for $N=3$.
We have discussed the cases of $N=2$ and $N=3$ in detail
showing how to combine the $I^{\alpha \beta, \gamma \delta}$
observables with the classification scheme in order to
identify any HF-symmetry,
regardless of the basis in which the Higgs potential may
be originally written in.
Our basis invariants $I^{\alpha \beta, \gamma \delta}$
may be applied to any other value of $N$,
by constructing the database of the classes of
symmetries possible for that
value of $N$.

This classification is also important because sometimes
one imposes a discrete symmetry only to find that the
potential becomes automatically invariant under a much larger
class of symmetries.
These may even be continuous,
implying the danger that Goldstone bosons
might appear.
We provide explicit examples of this problem.
This may even be relevant for studies
of the fermion sector.
For example,
Grimus \textit{et al.\/} discuss very general symmetry
realizations of texture zeros in the fermion sector
with the help of scalar fields onto which certain
discrete symmetries are imposed \cite{Gri}.
When using such techniques,
one must inspect also the Higgs potential in some detail,
including the symmetry breaking,
lest there be undesired massless scalars.

\begin{acknowledgments}
We are grateful to Owen Brison for useful discussions.
The work of P.\ M.\ F.\ is supported in part
by the Portuguese \textit{Funda\c{c}\~{a}o para
a Ci\^{e}ncia e a Tecnologia} (FCT) under contract
PTDC/FIS/70156/2006.
The work of J.\ P.\ S.\ is supported in part by FCT under
contract CFTP-Plurianual (U777).
\end{acknowledgments}

\appendix
\section{\label{app:matrices}Coupling structures for the
different classes of symmetries}

In this appendix we discuss the impact that the seven classes of
symmetries identified in the 3HDM have on the coupling
constants in the scalar potential.
We show the result in the special basis in which the symmetry
has one of the diagonal forms in Eq.~(\ref{seven_from_Z}).
In section~\ref{sec:fingerprints} we show how to turn this information
into a basis invariant fingerprint for the discrete symmetries.

The quadratic couplings distinguish three cases:
i) the most general potential,
where all entries of $Y_{ij}$ may be nonzero;
ii) the potential with one of the symmetries
$S_1$--$S_2$,
where the matrix $Y_{ij}$ is block diagonal and the
uppermost-leftmost $2 \times 2$ block is left
unconstrained;
and iii) the potential with one of the symmetries
$S_3$--$S_7$,
where the matrix $Y_{ij}$ is diagonal.
The first case corresponds to
3 real and 3 complex parameters
(for a sum of 9 real variables);
the second case corresponds to
3 real and 1 complex parameters
(5 real variables);
and the third case corresponds to
3 real parameters.

To see the impact on the quartic potential,
we organize the $Z_{ij,kl}$ tensor into
a matrix of matrices.
The uppermost-leftmost matrix corresponds to
the phases affecting $Z_{11,kl}$. 
The next matrix along the same line corresponds
to the phases affecting $Z_{12,kl}$, and so on...
We use the following notation for the various
entries
\begin{equation}
\left[
\begin{array}{ccc}
    \left[
    \begin{array}{ccc}
      r_1 & c_1 & c_2 \\
      c_1^\ast & r_4 & c_6 \\
      c_2^\ast & c_6^\ast & r_5 \\
    \end{array}
    \right]
&
    \left[
    \begin{array}{ccc}
      c_1 & c_3 & c_4 \\
      r_7 & c_7 & c_8 \\
      c_9^\ast & c_{12} & c_{13} \\
    \end{array}
    \right]
&
    \left[
    \begin{array}{ccc}
      c_2 & c_4 & c_5 \\
      c_9 & c_{10} & c_{11} \\
      r_8 & c_{14} & c_{15} \\
    \end{array}
    \right]
\\*[12mm]
    \left[
    \begin{array}{ccc}
      c_1^\ast & r_7 & c_9 \\
      c_3^\ast & c_7^\ast & c_{12}^\ast \\
      c_4^\ast & c_8^\ast & c_{13}^\ast \\
    \end{array}
    \right]
&
    \left[
    \begin{array}{ccc}
      r_4 & c_7 & c_{10} \\
      c_7^\ast & r_2 & c_{16} \\
      c_{10}^\ast & c_{16}^\ast & r_6 \\
    \end{array}
    \right]
&
    \left[
    \begin{array}{ccc}
      c_6 & c_8 & c_{11} \\
      c_{12}^\ast & c_{16} & c_{17} \\
      c_{14}^\ast & r_9 & c_{18} \\
    \end{array}
    \right]
\\*[12mm]
    \left[
    \begin{array}{ccc}
      c_2^\ast & c_9^\ast & r_8 \\
      c_4^\ast & c_{10}^\ast & c_{14}^\ast \\
      c_5 & c_{11}^\ast & c_{15}^\ast \\
    \end{array}
    \right]
&
    \left[
    \begin{array}{ccc}
      c_6^\ast & c_{12} & c_{14} \\
      c_8^\ast & c_{16}^\ast & r_9 \\
      c_{11}^\ast & c_{17}^\ast & c_{18}^\ast \\
    \end{array}
    \right]
&
    \left[
    \begin{array}{ccc}
      r_5 & c_{13} & c_{15} \\
      c_{13}^\ast & r_6 & c_{18} \\
      c_{15}^\ast & c_{18}^\ast & r_3 \\
     \end{array}
    \right]
\\
\end{array}
\right],
\label{SU3-Z_2}
\end{equation}
where $r_i$ ($i = 1 \dots 9$) are real and $c_i$ ($i = 1 \dots 18$) 
are complex.
In the basis of Eq.~(\ref{seven_from_Z}),
the symmetries $S_1$--$S_7$ set different combinations of
$c_i$ to zero but leave the real coefficients $r_i$
unconstrained.

The most general 3HDM has 9 real and 18 complex quartic
couplings for a total of 45 real variables.
Combining with the quadratic parameters,
we have 12 real and 21 complex parameters,
for a total of 54 real variables
(33 magnitudes and 21 phases).
However,
not all the variables have physical significance due to the
possibility of changing basis through any $3 \times 3$
unitary matrix
(which can be parametrized with 3 magnitudes,
5 relative phases, and 1 global phase).
Thus,
the most general 3HDM has 
30 magnitudes and 16 phases with
physical significance.

If the potential obeys the symmetry $S_1$,
in the basis of Eq.~(\ref{seven_from_Z})
the quartic couplings have the following structure,
\begin{equation}
\left[
\begin{array}{ccc}
    \left[
    \begin{array}{ccc}
      r_1 & c_1 & 0 \\
      c_1^\ast & r_4 & 0 \\
      0 & 0 & r_5 \\
    \end{array}
    \right]
&
    \left[
    \begin{array}{ccc}
      c_1 & c_3 & 0 \\
      r_7 & c_7 & 0 \\
      0 & 0 & c_{13} \\
    \end{array}
    \right]
&
    \left[
    \begin{array}{ccc}
      0 & 0 & c_5 \\
      0 & 0 & c_{11} \\
      r_8 & c_{14} & 0 \\
    \end{array}
    \right]
\\*[12mm]
    \left[
    \begin{array}{ccc}
      c_1^\ast & r_7 & 0 \\
      c_3^\ast & c_7^\ast & 0 \\
      0 & 0 & c_{13}^\ast \\
    \end{array}
    \right]
&
    \left[
    \begin{array}{ccc}
      r_4 & c_7 & 0 \\
      c_7^\ast & r_2 & 0 \\
      0 & 0 & r_6 \\
    \end{array}
    \right]
&
    \left[
    \begin{array}{ccc}
      0 & 0 & c_{11} \\
      0 & 0 & c_{17} \\
      c_{14}^\ast & r_9 & 0 \\
    \end{array}
    \right]
\\*[12mm]
    \left[
    \begin{array}{ccc}
      0 & 0 & r_8 \\
      0 & 0 & c_{14}^\ast \\
      c_5 & c_{11}^\ast & 0 \\
    \end{array}
    \right]
&
    \left[
    \begin{array}{ccc}
      0 & 0 & c_{14} \\
      0 & 0 & r_9 \\
      c_{11}^\ast & c_{17}^\ast & 0 \\
    \end{array}
    \right]
&
    \left[
    \begin{array}{ccc}
      r_5 & c_{13} & 0 \\
      c_{13}^\ast & r_6 & 0 \\
      0 & 0 & r_3 \\
     \end{array}
    \right]
\\
\end{array}
\right],
\label{quartic_S1}
\end{equation}
This corresponds to 9 (12) real and 8 (9) complex
parameters in the quartic couplings (in the scalar potential).

If the potential obeys the symmetry $S_2$,
in the basis of Eq.~(\ref{seven_from_Z})
the quartic couplings have the following structure,
\begin{equation}
\left[
\begin{array}{ccc}
    \left[
    \begin{array}{ccc}
      r_1 & c_1 & 0 \\
      c_1^\ast & r_4 & 0 \\
      0 & 0 & r_5 \\
    \end{array}
    \right]
&
    \left[
    \begin{array}{ccc}
      c_1 & c_3 & 0 \\
      r_7 & c_7 & 0 \\
      0 & 0 & c_{13} \\
    \end{array}
    \right]
&
    \left[
    \begin{array}{ccc}
      0 & 0 & 0 \\
      0 & 0 & 0 \\
      r_8 & c_{14} & 0 \\
    \end{array}
    \right]
\\*[12mm]
    \left[
    \begin{array}{ccc}
      c_1^\ast & r_7 & 0 \\
      c_3^\ast & c_7^\ast & 0 \\
      0 & 0 & c_{13}^\ast \\
    \end{array}
    \right]
&
    \left[
    \begin{array}{ccc}
      r_4 & c_7 & 0 \\
      c_7^\ast & r_2 & 0 \\
      0 & 0 & r_6 \\
    \end{array}
    \right]
&
    \left[
    \begin{array}{ccc}
      0 & 0 & 0 \\
      0 & 0 & 0 \\
      c_{14}^\ast & r_9 & 0 \\
    \end{array}
    \right]
\\*[12mm]
    \left[
    \begin{array}{ccc}
      0 & 0 & r_8 \\
      0 & 0 & c_{14}^\ast \\
      0 & 0 & 0 \\
    \end{array}
    \right]
&
    \left[
    \begin{array}{ccc}
      0 & 0 & c_{14} \\
      0 & 0 & r_9 \\
      0 & 0 & 0 \\
    \end{array}
    \right]
&
    \left[
    \begin{array}{ccc}
      r_5 & c_{13} & 0 \\
      c_{13}^\ast & r_6 & 0 \\
      0 & 0 & r_3 \\
     \end{array}
    \right]
\\
\end{array}
\right],
\label{quartic_S2}
\end{equation}
This corresponds to 9 (12) real and 5 (6) complex
parameters in the quartic couplings (in the scalar potential).

If the potential obeys the symmetry $S_3$,
in the basis of Eq.~(\ref{seven_from_Z})
the quartic couplings have the following structure,
\begin{equation}
\left[
\begin{array}{ccc}
    \left[
    \begin{array}{ccc}
      r_1 & 0 & 0 \\
      0 & r_4 & 0 \\
      0 & 0 & r_5 \\
    \end{array}
    \right]
&
    \left[
    \begin{array}{ccc}
      0 & 0 & c_4 \\
      r_7 & 0 & 0 \\
      0 & c_{12} & 0 \\
    \end{array}
    \right]
&
    \left[
    \begin{array}{ccc}
      0 & c_4 & 0 \\
      0 & 0 & c_{11} \\
      r_8 & 0 & 0 \\
    \end{array}
    \right]
\\*[12mm]
    \left[
    \begin{array}{ccc}
      0 & r_7 & 0 \\
      0 & 0 & c_{12}^\ast \\
      c_4^\ast & 0 & 0 \\
    \end{array}
    \right]
&
    \left[
    \begin{array}{ccc}
      r_4 & 0 & 0 \\
      0 & r_2 & 0 \\
      0 & 0 & r_6 \\
    \end{array}
    \right]
&
    \left[
    \begin{array}{ccc}
      0 & 0 & c_{11} \\
      c_{12}^\ast & 0 & 0 \\
      0 & r_9 & 0 \\
    \end{array}
    \right]
\\*[12mm]
    \left[
    \begin{array}{ccc}
      0 & 0 & r_8 \\
      c_4^\ast & 0 & 0 \\
      0 & c_{11}^\ast & 0 \\
    \end{array}
    \right]
&
    \left[
    \begin{array}{ccc}
      0 & c_{12} & 0 \\
      0 & 0 & r_9 \\
      c_{11}^\ast & 0 & 0 \\
    \end{array}
    \right]
&
    \left[
    \begin{array}{ccc}
      r_5 & 0 & 0 \\
      0 & r_6 & 0 \\
      0 & 0 & r_3 \\
     \end{array}
    \right]
\\
\end{array}
\right],
\label{quartic_S3}
\end{equation}
This corresponds to 9 (12) real and 3 (3) complex
parameters in the quartic couplings (in the scalar potential).

If the potential obeys the symmetry $S_4$,
in the basis of Eq.~(\ref{seven_from_Z})
the quartic couplings have the following structure,
\begin{equation}
\left[
\begin{array}{ccc}
    \left[
    \begin{array}{ccc}
      r_1 & 0 & 0 \\
      0 & r_4 & 0 \\
      0 & 0 & r_5 \\
    \end{array}
    \right]
&
    \left[
    \begin{array}{ccc}
      0 & 0 & c_4 \\
      r_7 & 0 & 0 \\
      0 & 0 & 0 \\
    \end{array}
    \right]
&
    \left[
    \begin{array}{ccc}
      0 & c_4 & 0 \\
      0 & 0 & 0 \\
      r_8 & 0 & 0 \\
    \end{array}
    \right]
\\*[12mm]
    \left[
    \begin{array}{ccc}
      0 & r_7 & 0 \\
      0 & 0 & 0 \\
      c_4^\ast & 0 & 0 \\
    \end{array}
    \right]
&
    \left[
    \begin{array}{ccc}
      r_4 & 0 & 0 \\
      0 & r_2 & 0 \\
      0 & 0 & r_6 \\
    \end{array}
    \right]
&
    \left[
    \begin{array}{ccc}
      0 & 0 & 0 \\
      0 & 0 & c_{17} \\
      0 & r_9 & 0 \\
    \end{array}
    \right]
\\*[12mm]
    \left[
    \begin{array}{ccc}
      0 & 0 & r_8 \\
      c_4^\ast & 0 & 0 \\
      0 & 0 & 0 \\
    \end{array}
    \right]
&
    \left[
    \begin{array}{ccc}
      0 & 0 & 0 \\
      0 & 0 & r_9 \\
      0 & c_{17}^\ast & 0 \\
    \end{array}
    \right]
&
    \left[
    \begin{array}{ccc}
      r_5 & 0 & 0 \\
      0 & r_6 & 0 \\
      0 & 0 & r_3 \\
     \end{array}
    \right]
\\
\end{array}
\right],
\label{quartic_S4}
\end{equation}
This corresponds to 9 (12) real and 2 (2) complex
parameters in the quartic couplings (in the scalar potential).

If the potential obeys the symmetry $S_5$,
in the basis of Eq.~(\ref{seven_from_Z})
the quartic couplings have the following structure,
\begin{equation}
\left[
\begin{array}{ccc}
    \left[
    \begin{array}{ccc}
      r_1 & 0 & 0 \\
      0 & r_4 & 0 \\
      0 & 0 & r_5 \\
    \end{array}
    \right]
&
    \left[
    \begin{array}{ccc}
      0 & 0 & c_4 \\
      r_7 & 0 & 0 \\
      0 & 0 & 0 \\
    \end{array}
    \right]
&
    \left[
    \begin{array}{ccc}
      0 & c_4 & 0 \\
      0 & 0 & 0 \\
      r_8 & 0 & 0 \\
    \end{array}
    \right]
\\*[12mm]
    \left[
    \begin{array}{ccc}
      0 & r_7 & 0 \\
      0 & 0 & 0 \\
      c_4^\ast & 0 & 0 \\
    \end{array}
    \right]
&
    \left[
    \begin{array}{ccc}
      r_4 & 0 & 0 \\
      0 & r_2 & 0 \\
      0 & 0 & r_6 \\
    \end{array}
    \right]
&
    \left[
    \begin{array}{ccc}
      0 & 0 & 0 \\
      0 & 0 & 0 \\
      0 & r_9 & 0 \\
    \end{array}
    \right]
\\*[12mm]
    \left[
    \begin{array}{ccc}
      0 & 0 & r_8 \\
      c_4^\ast & 0 & 0 \\
      0 & 0 & 0 \\
    \end{array}
    \right]
&
    \left[
    \begin{array}{ccc}
      0 & 0 & 0 \\
      0 & 0 & r_9 \\
      0 & 0 & 0 \\
    \end{array}
    \right]
&
    \left[
    \begin{array}{ccc}
      r_5 & 0 & 0 \\
      0 & r_6 & 0 \\
      0 & 0 & r_3 \\
     \end{array}
    \right]
\\
\end{array}
\right],
\label{quartic_S5}
\end{equation}
This corresponds to 9 (12) real and 1 (1) complex
parameters in the quartic couplings (in the scalar potential).
The single complex parameter appears in four entries.

If the potential obeys the symmetry $S_6$,
in the basis of Eq.~(\ref{seven_from_Z})
the quartic couplings have the following structure,
\begin{equation}
\left[
\begin{array}{ccc}
    \left[
    \begin{array}{ccc}
      r_1 & 0 & 0 \\
      0 & r_4 & 0 \\
      0 & 0 & r_5 \\
    \end{array}
    \right]
&
    \left[
    \begin{array}{ccc}
      0 & c_3 & 0 \\
      r_7 & 0 & 0 \\
      0 & 0 & 0 \\
    \end{array}
    \right]
&
    \left[
    \begin{array}{ccc}
      0 & 0 & 0 \\
      0 & 0 & 0 \\
      r_8 & 0 & 0 \\
    \end{array}
    \right]
\\*[12mm]
    \left[
    \begin{array}{ccc}
      0 & r_7 & 0 \\
      c_3^\ast & 0 & 0 \\
      0 & 0 & 0 \\
    \end{array}
    \right]
&
    \left[
    \begin{array}{ccc}
      r_4 & 0 & 0 \\
      0 & r_2 & 0 \\
      0 & 0 & r_6 \\
    \end{array}
    \right]
&
    \left[
    \begin{array}{ccc}
      0 & 0 & 0 \\
      0 & 0 & 0 \\
      0 & r_9 & 0 \\
    \end{array}
    \right]
\\*[12mm]
    \left[
    \begin{array}{ccc}
      0 & 0 & r_8 \\
      0 & 0 & 0 \\
      0 & 0 & 0 \\
    \end{array}
    \right]
&
    \left[
    \begin{array}{ccc}
      0 & 0 & 0 \\
      0 & 0 & r_9 \\
      0 & 0 & 0 \\
    \end{array}
    \right]
&
    \left[
    \begin{array}{ccc}
      r_5 & 0 & 0 \\
      0 & r_6 & 0 \\
      0 & 0 & r_3 \\
     \end{array}
    \right]
\\
\end{array}
\right],
\label{quartic_S6}
\end{equation}
This corresponds to 9 (12) real and 1 (1) complex
parameters in the quartic couplings (in the scalar potential).
Unlike what happened for $S_5$,
here the single complex parameter appears in two entries.

If the potential obeys the symmetry $S_7$,
in the basis of Eq.~(\ref{seven_from_Z})
the quartic couplings have the following structure,
\begin{equation}
\left[
\begin{array}{ccc}
    \left[
    \begin{array}{ccc}
      r_1 & 0 & 0 \\
      0 & r_4 & 0 \\
      0 & 0 & r_5 \\
    \end{array}
    \right]
&
    \left[
    \begin{array}{ccc}
      0 & 0 & 0 \\
      r_7 & 0 & 0 \\
      0 & 0 & 0 \\
    \end{array}
    \right]
&
    \left[
    \begin{array}{ccc}
      0 & 0 & 0 \\
      0 & 0 & 0 \\
      r_8 & 0 & 0 \\
    \end{array}
    \right]
\\*[12mm]
    \left[
    \begin{array}{ccc}
      0 & r_7 & 0 \\
      0 & 0 & 0 \\
      0 & 0 & 0 \\
    \end{array}
    \right]
&
    \left[
    \begin{array}{ccc}
      r_4 & 0 & 0 \\
      0 & r_2 & 0 \\
      0 & 0 & r_6 \\
    \end{array}
    \right]
&
    \left[
    \begin{array}{ccc}
      0 & 0 & 0 \\
      0 & 0 & 0 \\
      0 & r_9 & 0 \\
    \end{array}
    \right]
\\*[12mm]
    \left[
    \begin{array}{ccc}
      0 & 0 & r_8 \\
      0 & 0 & 0 \\
      0 & 0 & 0 \\
    \end{array}
    \right]
&
    \left[
    \begin{array}{ccc}
      0 & 0 & 0 \\
      0 & 0 & r_9 \\
      0 & 0 & 0 \\
    \end{array}
    \right]
&
    \left[
    \begin{array}{ccc}
      r_5 & 0 & 0 \\
      0 & r_6 & 0 \\
      0 & 0 & r_3 \\
     \end{array}
    \right]
\\
\end{array}
\right],
\label{quartic_S7}
\end{equation}
This corresponds to 9 (12) real 
parameters in the quartic couplings (in the scalar potential).
All complex parameters in the scalar potential vanish.

\end{document}